\documentclass[prl,superscriptaddress,twocolumn,showpacs]{revtex4-1}
\pdfoutput=1
\usepackage{amsmath, amsthm, amssymb, mathrsfs}
\usepackage[pdfborder={0 0 0}]{hyperref}

\usepackage{graphicx}
\usepackage{amsmath}
\usepackage{amssymb}
\usepackage{epstopdf}
\usepackage{grffile}
\usepackage{array}
\usepackage{tikz}
\usepackage[utf8]{inputenc}

\newcommand{\ie}{i.e.\ }
\newcommand{\etal}{\textit{et al.}\ }

\newcommand{\dd}{\mathrm{d}}
\renewcommand{\vec}[1]{\boldsymbol{#1}}

\begin{document}

\title{$p$-Wave Superfluidity by Spin-Nematic Fermi Surface Deformation}

\author{Jan Gukelberger}
\email[Corresponding author: ]{gukelberger@phys.ethz.ch}
\affiliation{Theoretical Physics, ETH Zurich, 8093 Zurich, Switzerland}

\author{Evgeny Kozik}
\affiliation{Physics Department, King's College, London WC2R 2LS, United Kingdom}

\author{Lode Pollet}
\affiliation{Department of Physics, Arnold Sommerfeld Center for Theoretical Physics and Center for NanoScience, University of Munich, Theresienstrasse 37, 80333 Munich, Germany}

\author{Nikolay Prokof'ev}
\affiliation{Department of Physics, University of Massachusetts, Amherst, MA 01003-4525, USA}

\author{Manfred Sigrist}
\affiliation{Theoretical Physics, ETH Zurich, 8093 Zurich, Switzerland}

\author{Boris Svistunov}
\affiliation{Department of Physics, University of Massachusetts, Amherst, MA 01003-4525, USA}

\author{Matthias Troyer}
\affiliation{Theoretical Physics, ETH Zurich, 8093 Zurich, Switzerland}

\date{\today}                                           

\begin{abstract}
We study attractively interacting fermions on a square lattice with dispersion relations exhibiting strong spin-dependent anisotropy. The resulting Fermi surface mismatch suppresses the $s$-wave BCS-type instability, clearing the way for unconventional types of order.
Unbiased sampling of the Feynman diagrammatic series using Diagrammatic Monte
Carlo methods reveals a rich phase diagram in the regime of intermediate coupling strength.
Instead of a proposed Cooper-pair Bose metal phase [A. E. Feiguin and M. P. A. Fisher, Phys. Rev. Lett. {\bf 103}, 025303 (2009)] we find an incommensurate density wave at strong anisotropy and two different $p$-wave superfluid states with unconventional symmetry
at intermediate anisotropy.
\end{abstract}


\maketitle

\begin{figure}
	\includegraphics[width=\columnwidth]{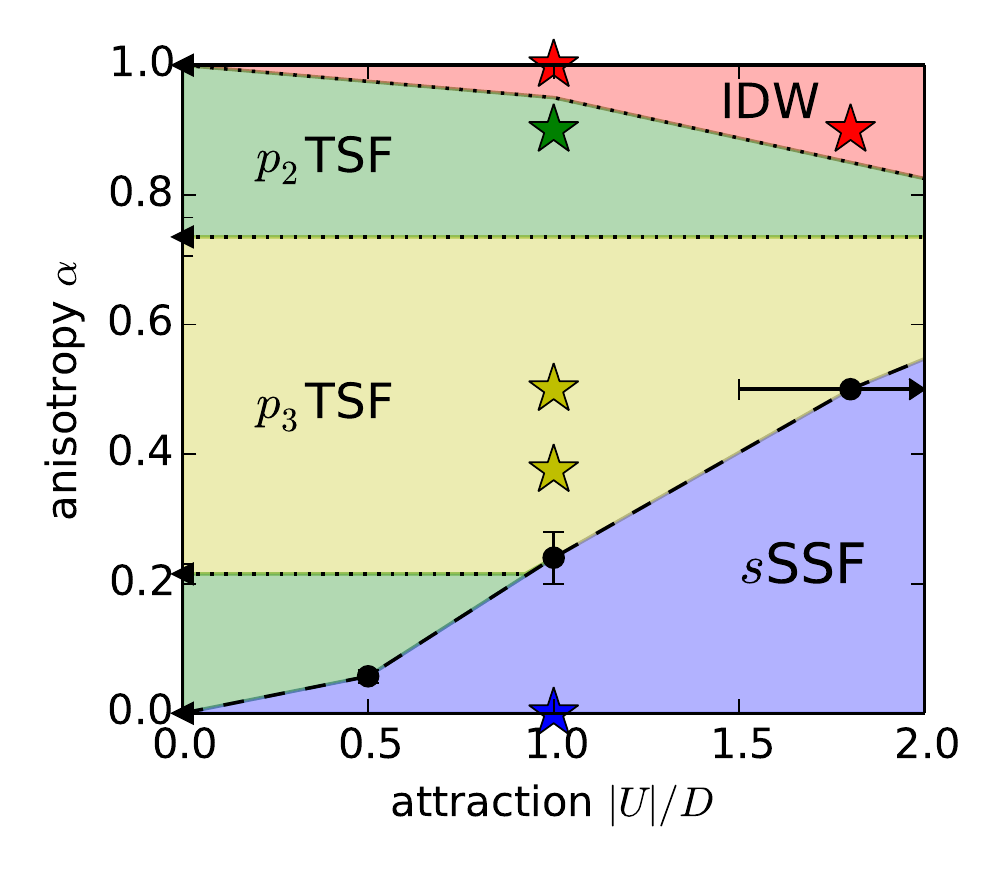}
	\caption{Schematic low temperature phase diagram at density $n=1.2$.
SSF/TSF are singlet/triplet superfluids, IDW an incommensurate density wave; see main text for an explanation of $s$/$p$ symmetry classifications.
Black triangles indicate phase transitions in the weak-coupling limit. Black circles mark the onset
of instability in the $s$SSF channel (with dashed lines interpolating between them).
Stars mark points where various ordering channels were compared to each other.
Colored regions delimited by dotted lines indicate possible extents of the phases consistent with our data. Except for the data points represented by symbols, we do not claim precise location of the respective phase boundaries.
At $\alpha \lesssim 0.2$ the $p_2$TSF and $p_3$TSF phases are nearly degenerate.}
\label{fig:phasediag}
\end{figure}


Usually a system of spin-$\frac{1}{2}$ fermions with an attractive effective interaction, even arbitrarily weak, is unstable towards a superfluid transition at low temperature where electrons at the Fermi surface (FS) form singlet pairs with zero center-of-mass momentum and the spectrum of single-particle excitations features a nodeless gap \cite{cooper1956bep,bardeen1957mts}.
More exotic types of order can arise when the $s$-wave superfluid state is suppressed by manipulating the spectrum or band fillings of the spin components, thereby creating a mismatch between the spin-up Fermi surface ($FS_{\uparrow}$) and the spin-down one ($FS_{\downarrow}$).
The latter option, {\it i.e.,} the introduction of spin population imbalance, may lead to the formation of pairs with finite total momentum and an inhomogeneous superfluid phase \cite{fulde1964sss,larkin1964iss}.
Another possibility is to keep the spin species equally populated but realize a spin-dependent FS deformation, which could either form spontaneously in a spin-nematic transition \cite{wu2007fli,raghu2009mtn,lee2009tum}
or be imposed externally when ultracold atoms are loaded into spin-dependent optical lattices \cite{feiguin2009eps}.

The latter scenario was suggested to harbor an exotic Cooper-pair Bose metal ground state, a putative metallic phase of tightly bound pairs with a gap for single-particle excitations, but no condensate and gapless bosonic excitations along so-called Bose surfaces in momentum space.
Until now a well-controlled investigation of this setup in two dimensions (2D) beyond the mean-field level, which in itself is debated~\cite{chiesa2014ce1,feiguin2014re1}, is lacking.
In the relevant regime of intermediate interaction strength this is challenging due to the absence of small parameters and the fact that broken spin-inversion symmetry causes a sign problem in determinant Monte Carlo simulations even for attractive interactions.
Density matrix renormalization group simulations on a ladder geometry see evidence for the existence of a one-dimensional analog of the Cooper-Pair Bose Metal \cite{feiguin2011epp} but extrapolations to the thermodynamic 2D limit are not straightforward.

In this Letter we solve the fundamental questions on the nature of the low temperature phases emerging once a spin-dependent anisotropy suppresses the conventional $s$-wave singlet superfluid by both a systematic study of the
weak-coupling limit and by Diagrammatic Monte Carlo (DiagMC) simulations at intermediate interaction strengths.
We find that mean-field calculations overestimate the stability of superfluid phases with trivial symmetry.
At low temperatures we find a rich phase diagram, shown in Fig.~\ref{fig:phasediag}, consisting of a conventional $s$-wave singlet superfluid ($s$SSF)
at weak anisotropy, an incommensurate density wave (IDW) at strong anisotropy and two different
$p$-wave triplet superfluids ($p_2$TSF and $p_3$TSF) at intermediate anisotropy.
Additionally, we clarify the mechanism leading to an indirect effective interaction between particles
with identical spins, enabling triplet pairing.

The model of Ref. \cite{feiguin2009eps} is a Hubbard-type Hamiltonian on a square lattice
\begin{align}
H = &
-\sum_{\substack{i, \sigma \\ \nu=\hat{x},\hat{y}}} \left( t_{\nu,\sigma} c_{i,\sigma}^\dag c_{i+\nu,\sigma} + h.c. \right) \nonumber\\&
+ U \sum_i n_{i,\uparrow} n_{i,\downarrow} - \mu \sum_{i,\sigma} n_{i,\sigma}
\label{eq:hamiltonian}
\end{align}
with spin-dependent anisotropic hopping amplitudes $t_{\nu,\sigma}$, on-site attraction $U < 0$, chemical potential $\mu$, and standard notations for on-site fermionic creation, $c_{i,\sigma}^\dag$,  and annihilation, $c_{i,\sigma}^{\,}$, operators with spin $\sigma=\uparrow,\downarrow$.
The hopping parameters are set to $t_{x\downarrow} = t_{y\uparrow} = t_a$, $t_{y\downarrow} = t_{x\uparrow} = t_b$
leading to an unpolarized system with balanced spin populations $\langle n_{i,\uparrow} \rangle = \langle n_{i,\downarrow} \rangle = n/2$.
All numerical results presented below are at fixed density $n=1.2$, which is equivalent to $n=0.8$ due to particle-hole symmetry. Other fillings will be discussed at the end of this Letter.
Without loss of generality we choose $t_b < t_a$ and define an anisotropy parameter $\alpha = 1 - t_b/t_a \in [0,1]$ so that $\alpha=0$ corresponds to the isotropic Hubbard model and $\alpha=1$ is the extreme anisotropy limit
where fermions can only move in one dimension
\footnote{Note that this definition of $\alpha$ differs from the $\alpha$ parameter chosen in Ref.~\onlinecite{feiguin2011epp}.}.
The half-bandwidth $D = 2(t_a+t_b)$ is chosen as the unit of energy.
%
The spin-dependent anisotropy breaks the continuous $SU(2)$ spin-rotation symmetry to a discrete $Z_2$ symmetry of combined spin inversion and space rotation by 90$^\circ$ and reduces the spatial symmetry to the point group of a rectangle, which has two irreducible representations with even and two with odd parity.
In the absence of spin rotation symmetry neither can the particle-particle channel be decomposed into singlet and triplet channels nor the particle-hole channel into density and spin channels. The terms ``singlet''/``triplet'' therefore refer to pairing between different/same spin species.
A ``density wave'' (DW) refers to an in-phase modulation of both spin densities.
In 2D, these phases have gapless Goldstone modes and exhibit algebraically decaying
order-parameter correlations instead of true long-range order at finite temperature \cite{herbut2007mac,gruner2009dws}.


\begin{figure}
	\includegraphics[width= \columnwidth]{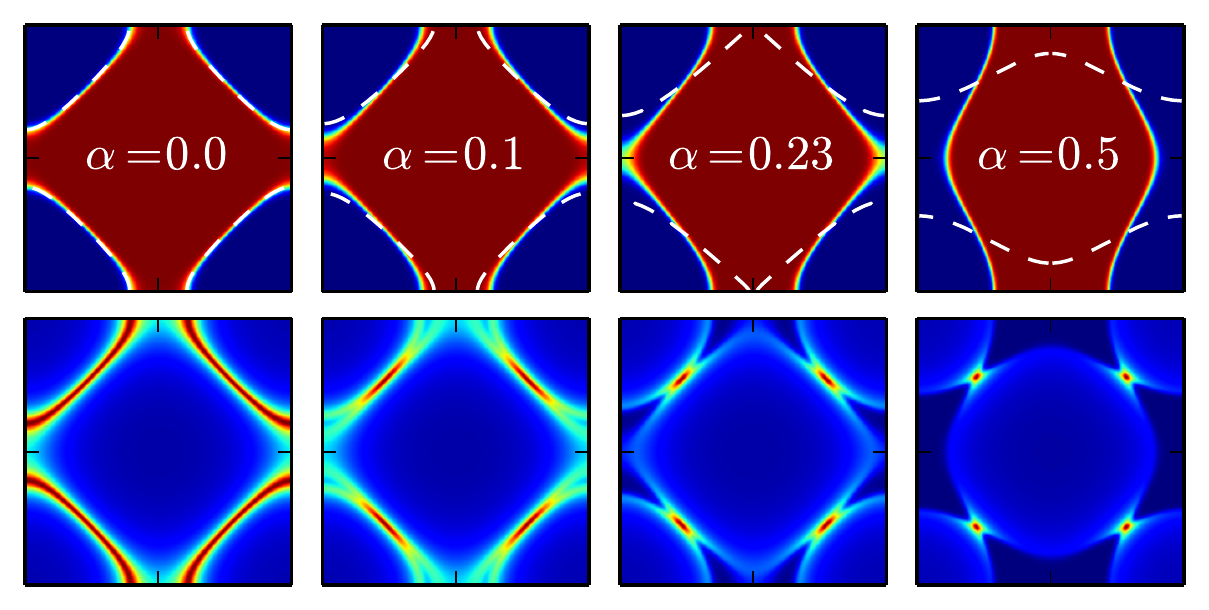}
	\caption{\textit{(top row:)} Momentum distributions (color plots) of free $\downarrow$-fermions
(dashed lines indicate $FS_{\uparrow}$) with hopping anisotropy $\alpha$ increasing from left to right at temperature $T/D=0.02$. Axes correspond to momenta $k_{x,y} \in [-\pi,\pi]$ in the first Brillouin zone of the square lattice.
	At $\alpha^* \approx 0.23$ the FS topology changes from a single closed contour (``2D like'') to two disconnected lines that wind around the BZ boundaries in one direction (``1D like'').
	\textit{(bottom row:)} Pair propagator $\chi^{pp}_{\uparrow\downarrow}$ for the same systems at zero frequency. The maximum value $D/4T$ (red color) is independent of $\alpha$ and diverges linearly with inverse temperature. However,
for $\alpha \neq 0$ its support shrinks to four discrete points in the $T \rightarrow 0$ limit,
rendering the integral over ${\bf k}$ finite.
	}
	\label{fig:n0k}
\end{figure}

As all the phases we find in our simulations can be understood from a weak-coupling perspective we first analyze the system in the $|U| \to 0$ limit before presenting DiagMC results for finite $U$.
Our general approach is to look for instabilities of the Fermi liquid when lowering the temperature.
A phase transition is signalled by the divergence of a correlation function and hence of the two-particle vertex $\tilde{\Gamma}$, which is related 
by the Bethe-Salpeter equation \cite{fetter2003quantum}
\tikzstyle{bline}=[baseline={([yshift=-0.6ex]current bounding box.center)}] 
\tikzstyle{vertex}=[rectangle,draw=black]
\begin{align}
\tikz[bline]{ \node[vertex]{$\tilde{\Gamma}$}; }
&=
\tikz[bline]{ \node[vertex]{$\Gamma^x$}; }
+
\tikz[bline]{ \node[vertex]{$\Gamma^x$}; }
\tikz[bline]{ \node[inner sep = 1pt,circle,draw]{$\chi^x$}; }
\tikz[bline]{ \node[vertex]{$\tilde{\Gamma}$}; }
\end{align}
to the particle-particle ($x=pp$) or particle-hole ($x=ph$) irreducible vertex $\Gamma$ and a product of two single-particle propagators $\chi = G G$.
$\tilde{\Gamma}$ diverges when the largest eigenvalue of the kernel $(-\Gamma \chi)$ reaches unity.
Decomposing the vertices according to spin and spatial symmetry and monitoring their leading eigenvalues we can hence detect and characterize phase transitions into  ordered states.

%
%
In the isotropic model the dominant weak-coupling instability
is the formation of Cooper-pair singlets with zero center-of-mass momentum and $s$-wave symmetry ($s$SSF). 
To first order in the interaction $\Gamma = U$ such that the leading eigenvalue
\footnote{We use the shorthand notation $k \equiv (i \omega_n,\vec{k})$ and $\int \dd k \equiv \int_{BZ} \frac{\dd^2 k}{(2 \pi)^2} T \sum_n$ for the Brillouin zone integral and sum over Matsubara frequencies.}
\begin{align} \label{eq:chiQ}
       \lambda^{(1)}_{sSSF}= -U \int \dd k \, G_\uparrow(k) G_\downarrow(-k)
\end{align}
will equal unity at a finite temperature for any $U < 0$ because the integral over the pair propagator
diverges logarithmically with decreasing temperature.
A finite anisotropy $\alpha \neq 0$ reduces the overlap
between $FS_{\uparrow}$ and $FS_{\downarrow}$ to four discrete crossing points, as illustrated in Fig.~\ref{fig:n0k}. This renders the integral \eqref{eq:chiQ} finite in the $T \to 0$ limit,
removing the weak-coupling Cooper instability.
In the extreme anisotropy limit $\alpha=1$ the Fermi edges are straight lines such that the wave vector $\vec{Q}=(2k_F,2k_F)$ provides perfect nesting conditions for the particle-hole propagators of both spin species at arbitrary filling, leading to a well-known 1D instability \cite{gruner2009dws}.
The result is an IDW along the lattice diagonals.
At generic anisotropy $\alpha < 1$ the weak-coupling instability in the
particle-hole propagator remains only at half filling 
where nesting at the staggered wave-vector $\vec{Q}=(\pi,\pi)$ is expected
to lead to checkerboard density order.


\begin{figure}
	\includegraphics[width=\columnwidth]{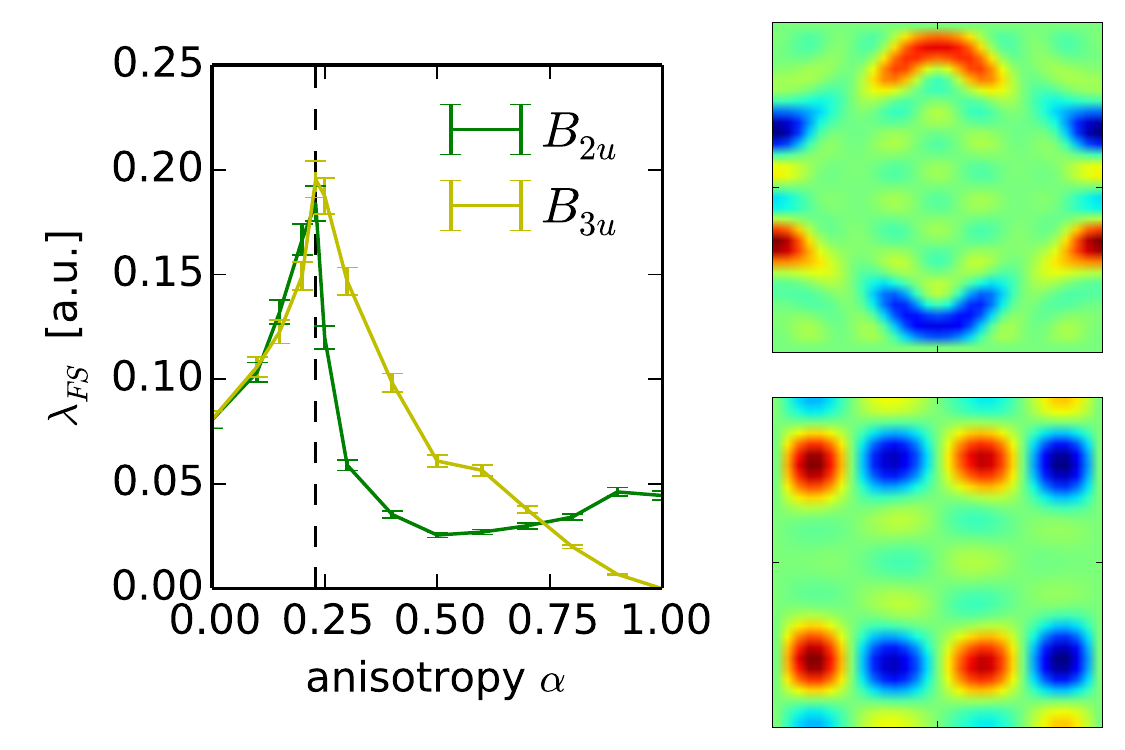}
	\caption{\textit{(left:)} Weak-coupling eigenvalues for pairing of $\uparrow$ particles. The dashed vertical line marks the crossover at $\alpha^*$ from 2D to 1D topology of the FS.
	\textit{(right:)} Momentum-space structure of the leading pairing eigenvector obtained with DiagMC at finite attraction $|U|=D$. Axes are the same as in Fig.~\ref{fig:n0k}. As in the weak-coupling analysis the leading instability at intermediate anisotropy $\alpha=0.375$ belongs to the representation $B_{3u}$ with horizontal nodal line $k_y=0$ \textit{(top)} whereas at large anisotropy $\alpha=0.9$ the $B_{2u}$ configuration with vertical node $k_x=0$ dominates \textit{(bottom)}.}
	\label{fig:weakgammappuu}
\end{figure}

Away from these special lines in the $\alpha - n$ phase diagram there are no instabilities to first order in $U$ as there is no direct interaction between identical particles. But at second order the particle-hole bubble
\begin{equation} \label{eq:gammappuu}
	\Gamma^{pp}_{\uparrow\uparrow}(k-k') = U^2 \int \dd k_1 \, G_\downarrow(k_1) G_\downarrow(k_1+k-k')
\end{equation}
mediates an effective interaction between the same-spin particles. As $FS_{\sigma}$ trivially matches with itself
there is a generic superfluid instability in the triplet channel, which becomes relevant whenever all other instabilities are removed. Due to fermionic antisymmetry the triplet pairs have odd parity (``$p$-wave'')
\footnote{Channels with odd frequency symmetry are irrelevant in the weak-coupling limit as the effective vertex vanishes at $\omega \to 0$, cancelling the  pair propagator divergence.}.
The point group's two odd irreducible representations are $B_{2u}$ and $B_{3u}$, which differ in the position of the nodal line but are related to each other by a 90$^\circ$ rotation such that they merge into the 2D representation of the square lattice in the isotropic limit.
Numeric calculations of $(-\Gamma \chi)$ with the second-order vertex \eqref{eq:gammappuu} show logarithmically diverging eigenvalues in both sectors. The prefactors of the $\ln T$ terms strongly depend on anisotropy (and filling) as shown in Fig.~\ref{fig:weakgammappuu}.
While intermediate anisotropy in general favors the configuration where the nodal line is parallel to the FS patches ($B_{3u}$ for $\uparrow$ spins, upper right panel in Fig.~\ref{fig:weakgammappuu}) the pairing vertex in this sector vanishes at extreme anisotropy because the $\downarrow$ spins mediating the effective interaction can only move in the direction of the nodal line, leaving only the configuration with the nodal line cutting through the FS of the pair's constituents ($B_{2u}$, lower right).
For $\alpha \to 0$ both configurations become degenerate.
As the model is invariant under a combined spin inversion and 90$^\circ$ rotation of space, both species reach the superfluid transition at the same temperature but in different symmetry sectors. To leading order,
the resulting state consists of independent condensates for the $\uparrow$ and $\downarrow$ spins, with the order parameter of one condensate having a horizontal nodal line and the other a vertical one. Still, the effective interaction holding the pairs together is purely due to the other species.
We refer to the superfluid phase where the $\uparrow \uparrow$ pairs have $B_{2u}$ or $B_{3u}$ symmetry as $p_2$TSF and $p_3$TSF, respectively.


\begin{figure}
	\includegraphics[width= \columnwidth]{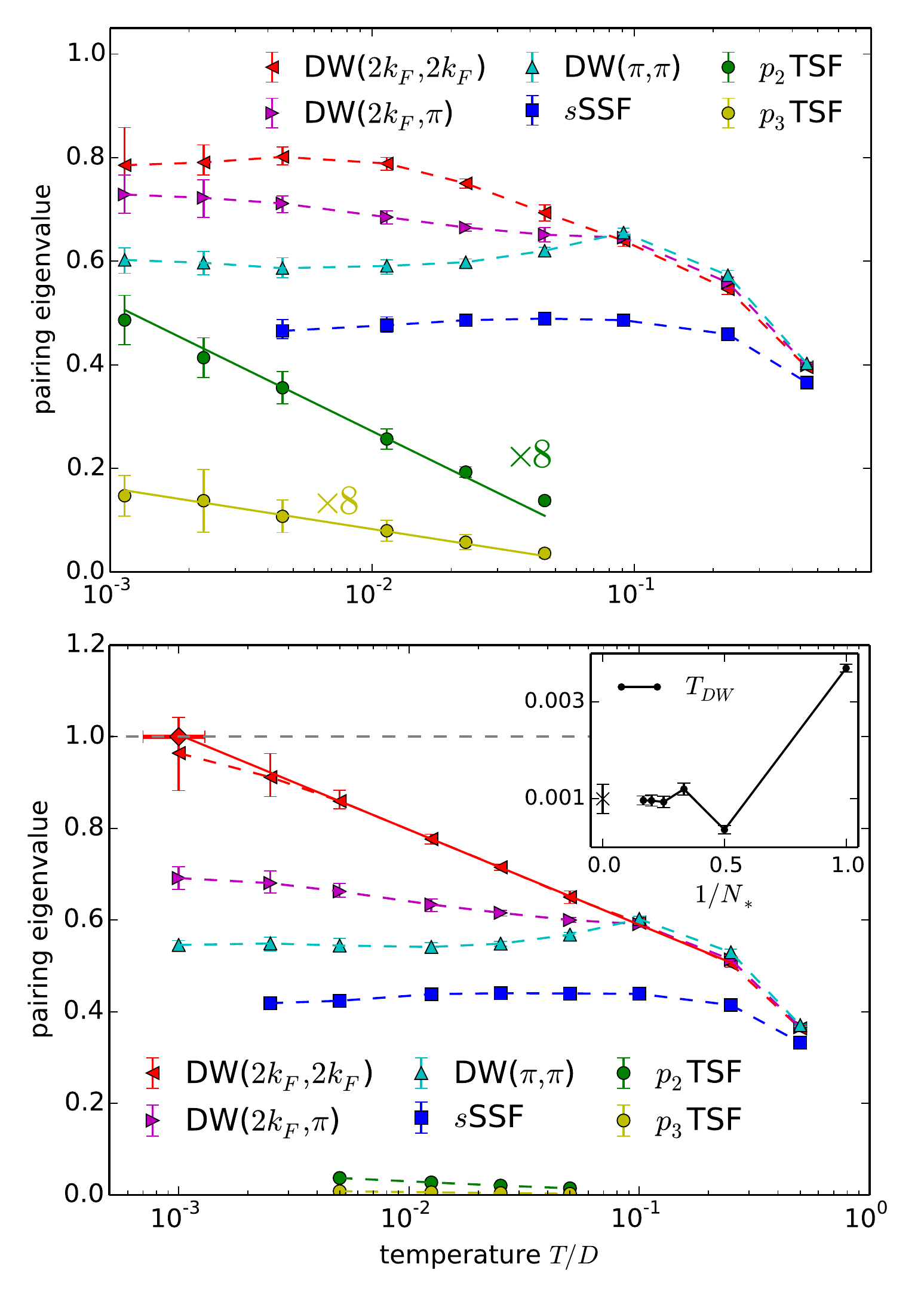}
	\caption{Temperature dependence of the leading eigenvalues at $|U|=D$ for strong anisotropy $\alpha=0.9$ \textit{(top)} and $\alpha=1$ \textit{(bottom)}. Dashed lines are guides to the eye. The solid lines going through the $p$-wave data points show the $T$-dependence predicted by Fermi-liquid theory based on Fermi-liquid parameters and eigenvalues
of the irreducible pairing vertex on the FS with $T \to 0$ extrapolation; only a constant offset accounting for high-temperature effects has been fitted to the finite $T$ data points. The solid line going through the $Q=(2k_F,2k_F)$ eigenvalues is a linear fit in $\ln T$. The inset shows convergence of the transition
temperature $T_{DW}$ with diagram order $N_*$.
The error bars on DW and $s$SSF data points at the lowest temperatures are systematic and dominated by
extrapolation in the number of Matsubara frequencies.}
	\label{fig:lambda_T_U1_alpha0.9}
\end{figure}

In order to confirm that this weak-coupling picture holds at finite $U$ we turn to DiagMC simulations,
which sample the bare many-body Feynman diagrammatic series directly in the thermodynamic limit  \cite{prokofev2007,VanHoucke2010,kozik2010}.
Since the diagrammatic series have a sign that vanishes exponentially in the diagram order the sampling process needs to be restricted to diagrams with order $N < N_*$.
By varying the cut-off, $N_*$, convergence of the results is checked, rendering the DiagMC approach well-controlled.
Whenever convergence with the diagram order is achieved, the result in the considered regime is expected to be exact.
In previous work the method has been thoroughly checked against other numerical techniques \cite{kozik2010,kozik2014}
and experiments \cite{VanHoucke2012}.
We find that all phases identified in the weak-coupling analysis extend to finite $U$.
As expected, the lowest-order $s$SSF and IDW instabilities
survive to finite anisotropy. Still, the FS mismatch is remarkably efficient in suppressing
these instabilities already  at
$|U| \approx D$, leaving a large domain in the phase diagram of Fig.~\ref{fig:phasediag} where the conventional eigenvalues saturate below unity at low temperature, as in the upper panel of Fig.~\ref{fig:lambda_T_U1_alpha0.9}, and only $p$-wave
order is present. Points where the saturated $s$SSF eigenvalue drops below unity are marked by black circles in Fig.~\ref{fig:phasediag}.
Transition temperatures for the unconventional superfluids are exponentially low
for the range of interactions we can access with our method so we cannot track their
eigenvalues down to temperatures close to $T_c$ as we did for the other channels.
Nevertheless, we are confident that these channels will ultimately diverge for two reasons:
First, we clearly observe the eigenvalues in other channels
saturate at low temperature when the Fermi edges are sharp enough to resolve
the anisotropy-caused mismatch. This leaves only the diverging $p$-wave SF channels.
Second, we observe the self-energy and the irreducible vertex converge at low temperature.
By extracting Fermi-liquid parameters and pairing eigenvalues on the FS from $T \to 0$ extrapolations
we obtain the asymptotic strength of the $\ln T$ divergence predicted by BCS theory. These
predictions match our finite temperature data remarkably well, verifying that we are indeed
observing the asymptotic low-temperature behavior \footnote{See appendix for a derivation of the pairing eigenvalues' asymptotic low-temperature behavior and additional data characterizing the phase diagram.}.

\begin{figure}
	\includegraphics[width=\columnwidth]{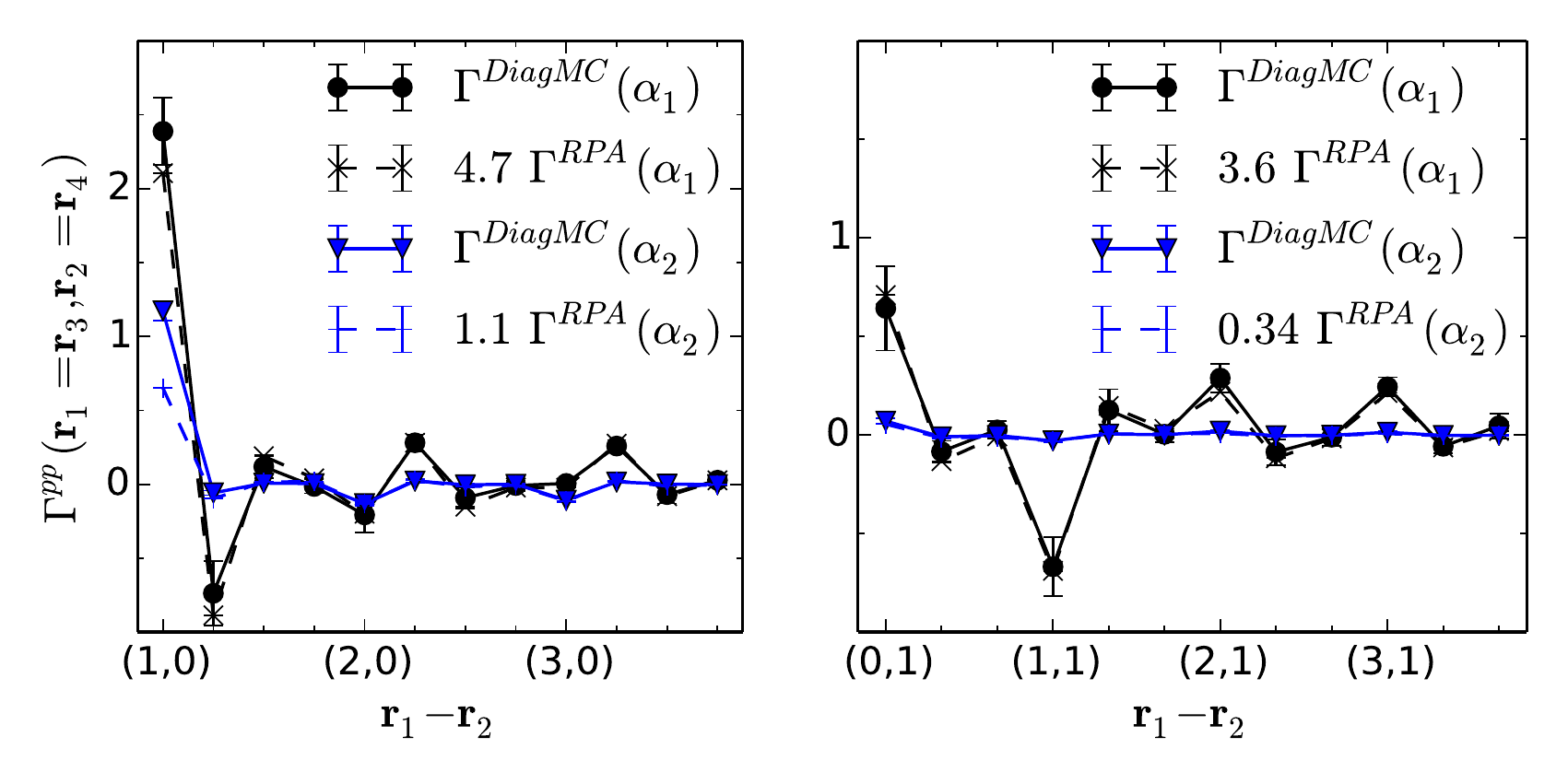}
	\caption{Spatial structure of the irreducible vertex in the $p_2$TSF  \textit{(left)} and $p_3$TSF \textit{(right)} channels for the anisotropies $\alpha_{1}=0.375$ and $\alpha_2=0.9$ calculated by DiagMC (for $|U|=D$) and within RPA approximation.
RPA data has been calculated for weak interaction $|U|=1.8 t_a$ and scaled with a constant factor
for each channel.}
	\label{fig:gammarpa}
\end{figure}

Having established the existence of the triplet superfluids $p_2$TSF and $p_3$TSF in spite of the absence of any direct interaction, we now investigate the mechanism that mediates an effective interaction between identical spins. Going a step beyond the weak coupling analysis, where the interaction is through a virtual particle-hole pair, we calculate the irreducible pairing vertex in RPA approximation \cite{anderson2008basic}.
Quantitatively accurate results cannot be expected from RPA, mainly due to the complete neglect of quasiparticle renormalizations of propagators and interactions. In fact, the RPA expression diverges for larger $|U| \sim D$, corresponding to a significant overestimation of the regime with a DW instability. 
But by performing the calculation at a reduced interaction and scaling the resulting vertex by a constant factor the spatial structure of the exact vertex 
is reproduced extremely well and over a wide range of anisotropy, see Fig.~\ref{fig:gammarpa}. 
As it seems unlikely that processes of different nature and hence diagrammatic structure would lead to exactly the same spatial structure, we conclude that the pairing interaction responsible for both triplet superfluids is predominantly mediated by density fluctuations.


The limited data we collected closer to half filling than the density $n=1.2$ chosen above look qualitatively similar; in general we expect the DW($\pi,\pi$) order to become more important the closer the system is to the $n=1$ point.
Farther away from half filling there are indications of a jump in $n(\mu)$, signalling phase separation towards a completely filled system, which is also observed in world-line Monte Carlo simulations for the system at full anisotropy and strong interaction \cite{lodeworms}.
Generally, we cannot exclude that our diagrammatic approach may miss  or underestimate signs of phase separation.
Due to slow convergence of DiagMC results and strong finite size effects and long autocorrelation times in world-line Monte Carlo we leave this question for future work.

The Cooper-pair Bose metal phase was proposed to exist at strong anisotropy $|1-\alpha| \ll 1$ and interaction $|U| \sim 2D$ \cite{feiguin2009eps} where our results indicate a density-ordered ground state. While both the IDW and the Bose metal are expected to have algebraically decaying density correlations with singular features at $Q=(2k_F,2k_F)$ the IDW has divergent correlations at finite temperature and true long-range order in the ground state whereas the Bose metal correlations remain finite and not ordered even at zero temperature.
We have presented clear evidence for the divergence of the $DW(2k_F,2k_F)$ density correlator, implying the presence of a density-ordered ground state at strong anisotropy from weak to intermediate interactions.
In the regime of strong interactions $|U| \gtrsim 2D$ we cannot reliably determine the exact nature of the low temperature phase due to slow convergence of the diagrammatic series but only ascertain a close competition between density waves with different wave vectors. It would be rather surprising if strong interactions replaced the ordered ground state with a metallic one.


\begin{acknowledgments}
We acknowledge useful discussions with M.~Dolfi, A.~Feiguin, M.~Fisher, M.~Iazzi and L.~Wang and thank A.\ Feiguin for his comments on the manuscript.
This work was supported by the Swiss National Science Foundation through the National Competence Center in Research QSIT, the National Science Foundation under Grant No.\ PHY-1314735, FP/ERC Starting Grant No.\ 306897, European Research Council (ERC) Advanced Grant SIMCOFE, and the MURI Program ``New Quantum Phases of Matter'' from the Air Force Office of Scientific Research (AFOSR). M.T.\ acknowledges the hospitality of the Aspen Center for Physics, supported by National Science Foundation (NSF) Grant No.\ 1066293.
We used the ALPS libraries for simulations and data evaluation \cite{alps20,alps13}.
Simulations were performed on the M\"onch and Brutus clusters of ETH Zurich.
\end{acknowledgments}


\appendix*
\section{Appendix}

\subsection{Asymptotic triplet-SF eigenvalues}

\begin{figure}
	\includegraphics[width= \columnwidth]{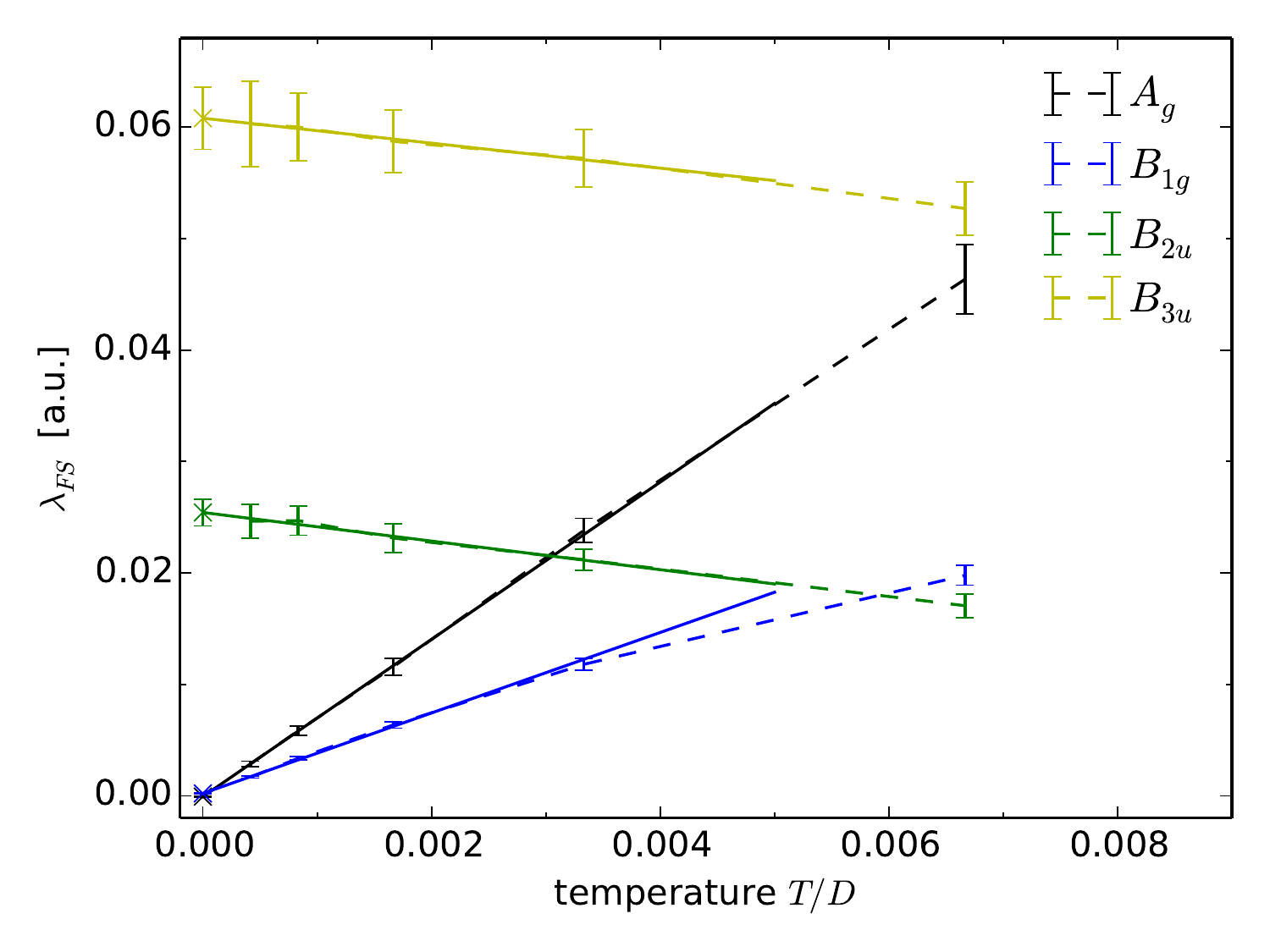}
	\caption{Extrapolation of the leading eigenvalues of the Bethe-Salpeter kernel \eqref{eq:tripletbsk} restricted to the lowest Matsubara frequencies $\omega=\omega'=i \pi T$ to zero temperature to determine the slope of
the $\ln T$-divergence in the weak-coupling limit.
	Shown is the data for intermediate anisotropy $\alpha=0.5$ at $n=1.2$. Dashed lines are guides to the eye, solid lines are linear fits, crosses at $T=0$ mark the extrapolated values. While eigenvalues for all four irreducible representations of the point group $D_{2h}$ are displayed, the even representations have odd frequency symmetry and hence vanish at $\omega \to 0$.
	The extrapolated values of odd representations are plotted versus anisotropy in Fig.~\ref{fig:weakgammappuu}.}
	\label{fig:weaklambda_T}
\end{figure}

The generic form of the Bethe-Salpeter kernel for superfluidity with zero center-of-mass momentum is
\begin{align} \label{eq:tripletbsk}
K(\omega_n,\vec{k}|\omega_{n'},\vec{k}') = -\frac{T}{(2\pi)^2} \chi(\omega_n,\vec{k}) \Gamma(\omega_n,\vec{k}|\omega_{n'},\vec{k}')
\end{align}
with pair propagator
$\chi(\omega_n,\vec{k}) = G(\omega_n,\vec{k}) G(-\omega_n,-\vec{k})$
and two-particle irreducible vertex $\Gamma$.
At low temperature $T \ll T_F$ it is dominated by processes taking place on
the FS and at vanishing frequency where the pair propagator diverges whereas the irreducible vertex converges to a smooth zero-temperature value such that its arguments can be restricted to the FS ($k,k' \to k_F$) and $\omega_n, \omega_{n'} \to 0$.
After this replacement only the pair propagator dependends on frequency $\omega_n$ and on the deviation from the FS $k-k_F$. Both variables are readily integrated out with standard techniques, linearizing the dispersion around the FS. \cite{abrikosov1975mqf} 
Then, discretizing the FS into segments $s$ of length $l_s$, the asymptotic scaling of the Bethe-Salpeter eigenvalue with temperature is
\begin{align} \label{eq:asymptlambda}
\lambda \propto - \lambda_{FS} \ln \frac{T}{T_F} \,,
\end{align}
where $\lambda_{FS}$ is determined by a discrete eigenvalue problem
\begin{align} \label{eq:lambdafs}
\lambda_{FS} = \mathrm{eig}\left( - \frac{l_s Z_s^2}{(2 \pi)^2 v_{F,s}} \Gamma_{s,s'} \right)  \,,
\end{align}
with $Z_s$ and $v_{F,s}$ the quasiparticle weight and Fermi velocity, respectively, at momentum $\vec{k}_{F,s}$.
In practice we split the FS into $>200$ segments $s$, making the discretization error negligible.
\begin{figure}
	\includegraphics[width= \columnwidth]{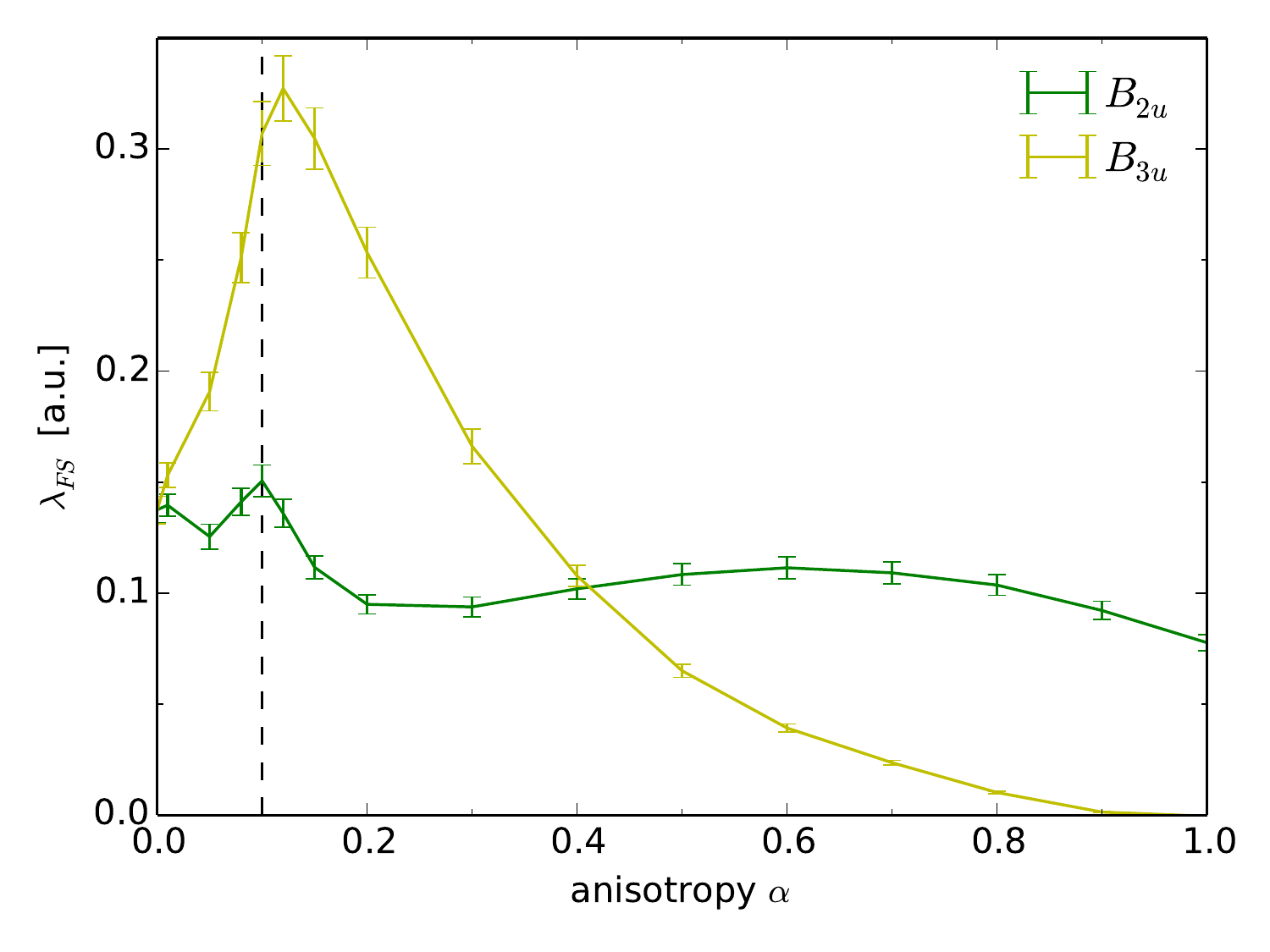}
	\caption{Weak coupling eigenvalues for pairing of $\uparrow$-particles at density $n=1.1$. The dashed vertical line marks the FS topology change $\alpha^*$. The fact that here the anisotropy quickly splits the $B_{2u}$ and $B_{3u}$ representations suggests that the near-degeneracy in a wide regime $0 < \alpha \lesssim 0.2$ at density n=1.2 (Fig.~\ref{fig:weakgammappuu}) is accidental.}
	\label{fig:weakgammappuu_n0.55}
\end{figure}
In the $|U| \to 0$ limit the leading instability is solely determined by the largest $\ln T$ prefactor.
Computing $\lambda_{FS}$ in each channel from the leading order diagram and extrapolating to $T \to 0$ (Fig.~\ref{fig:weaklambda_T}) therefore directly yields the weak-coupling phase diagram as shown in Fig.~\ref{fig:weakgammappuu} and Fig.~\ref{fig:weakgammappuu_n0.55} for different fillings.
At finite interaction the position of the FS $\vec{k}_{F,s}$, the quasiparticle weight $Z_s$ and Fermi velocity $v_{F,s}$ are extracted from the proper self-energy and multiplied with the irreducible vertex evaluated on the FS $\Gamma_{s,s'}=\Gamma(\omega_0,\vec{k}_{F,s}|\omega_{0}, \vec{k}_{F,s'})$. 
Repeating this procedure with data for different temperatures we obtain $\lambda_{FS}(T)$, which is then extrapolated to $T \to 0$.
In contrast to the weak-coupling case there are contributions from processes with higher energy, which will freeze out at low temperatures and hence not contribute to the asymptotic scaling. They are accounted for by fitting a constant offset to the temperature dependence of the full eigenvalues calculated by DiagMC.
Still, agreement between finite temperature data and the asymptotic scaling form \eqref{eq:asymptlambda} with a zero-temperature extrapolation of $\lambda_{FS}$ is a non-trivial check that an apparent $\ln T$ dependence is indeed the asymptotic $T \to 0$ behavior.

\subsection{Additional DiagMC data}

\begin{figure}
	\includegraphics[width= \columnwidth]{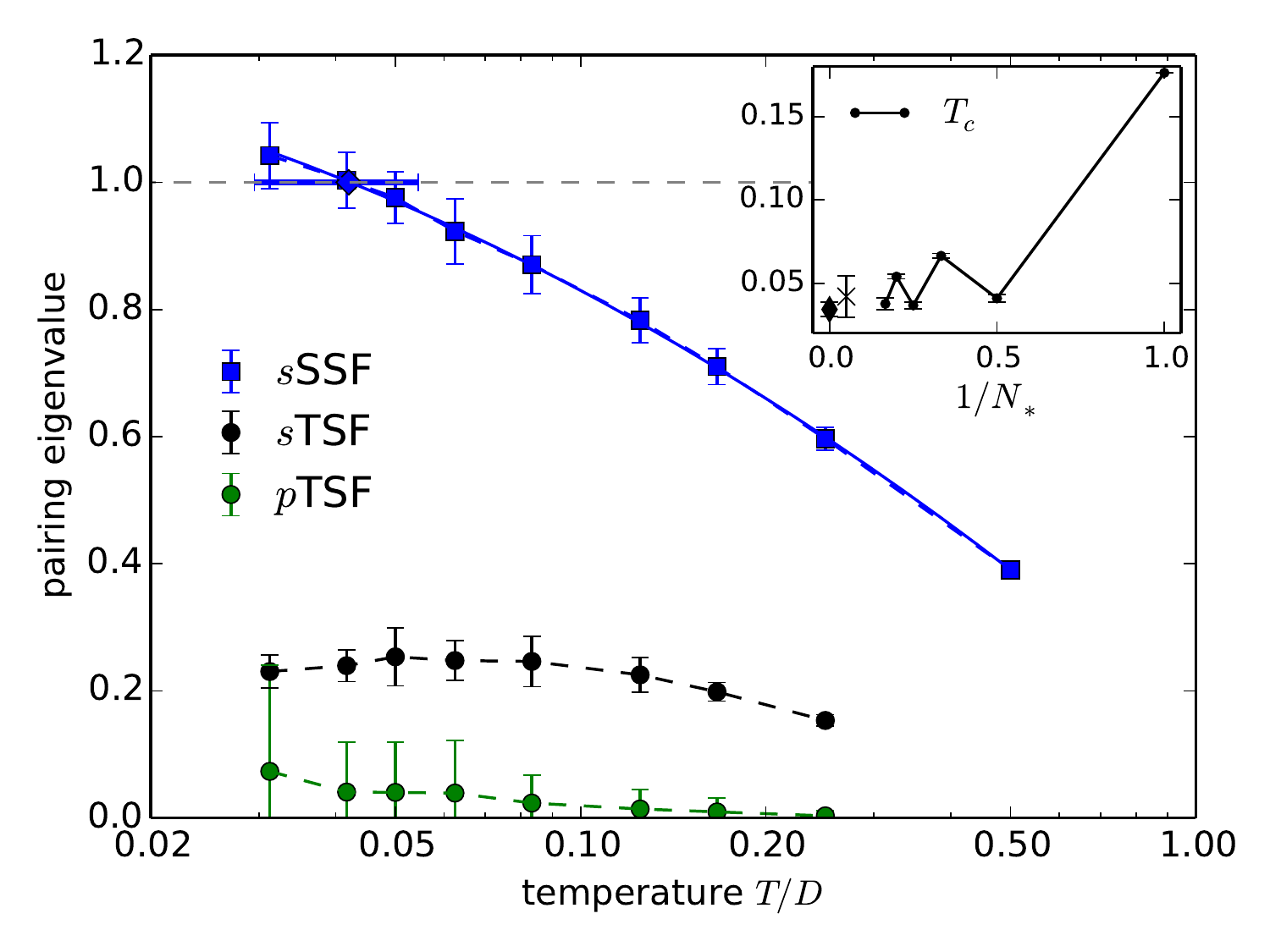}
	\caption{Superconducting eigenvalues for the isotropic model at $|U|=D$. Dashed lines are guides to the eye, the solid line is a second order polynomial fit in $\ln T$ used to determine the transition temperature $T_c$ where the $s$-wave singlet eigenvalue crosses unity. In the triplet channel the (odd-frequency) $s$-wave sector dominates over the $p$-wave one at high temperature but saturates at low $T$ due do the node at $\omega=0$. \textit{(Inset:)} Separately doing the fit for each cut-off order $N_*$ we obtain the dependence of $T_c$ on $N_*$, which we extrapolate to $T_c/D=0.042(12)$ (black cross), consistent with DetMC results for similar densities in Ref.~\onlinecite{paiva2004}, which we interpolate to $T_c/D \approx 0.034(4)$ (black diamond). }
	\label{fig:lambda_T_U1_alpha0}
\end{figure}

\begin{figure}
	\includegraphics[width=\columnwidth]{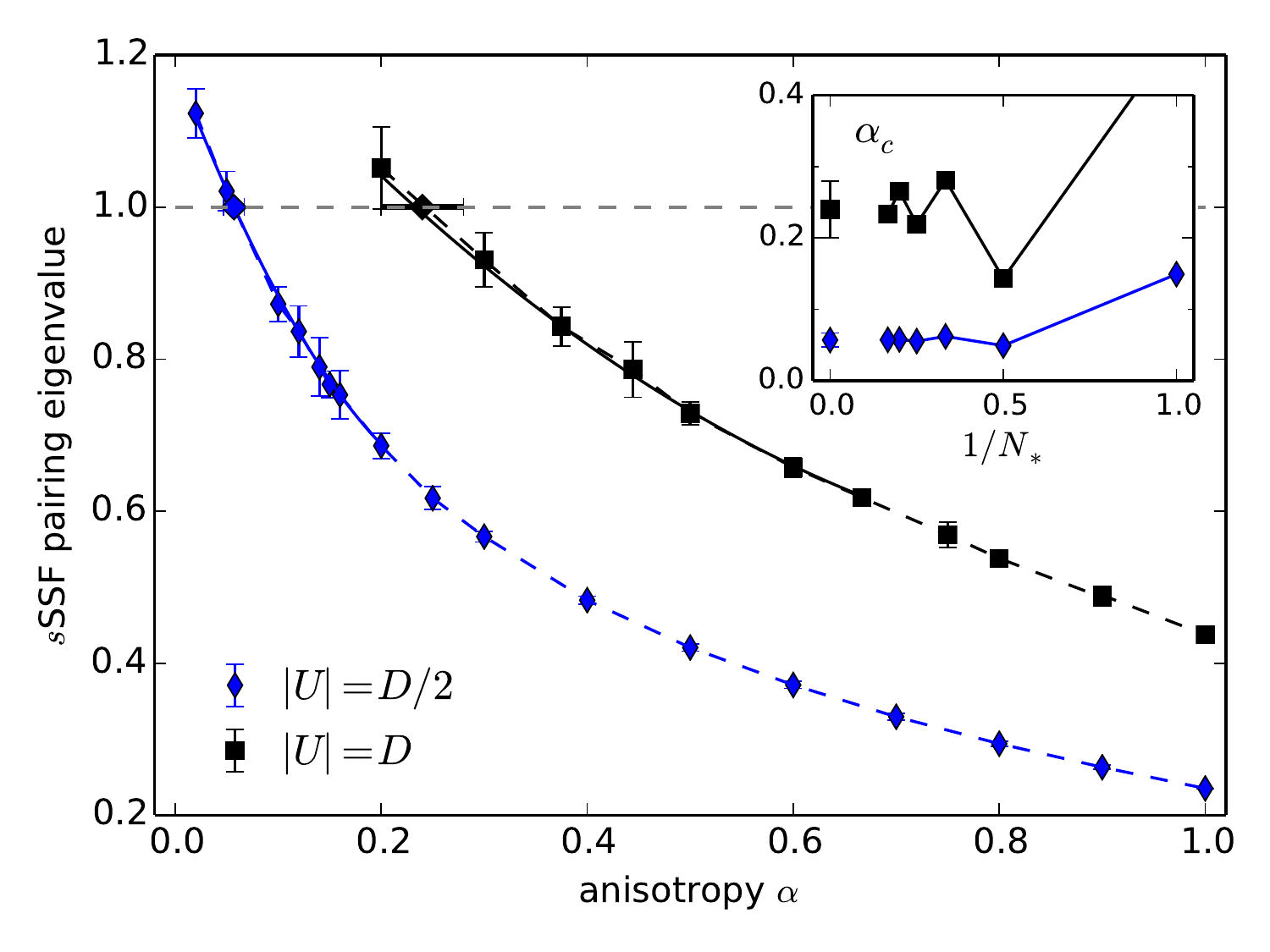}
	\caption{Saturated eigenvalues in the $s$SSF channel at low temperature for constant interaction $|U|$.
	We find critical anisotropies $\alpha_c(D/2) = 0.057(10)$ and $\alpha_c(D) = 0.24(4)$ above which there is no BCS instability at any temperature. The inset shows $\alpha_c$ extrapolations in the cut-off $N_*$.}
	\label{fig:swave-onset-alpha}
\end{figure}
\begin{figure}
	\includegraphics[width=.96 \columnwidth]{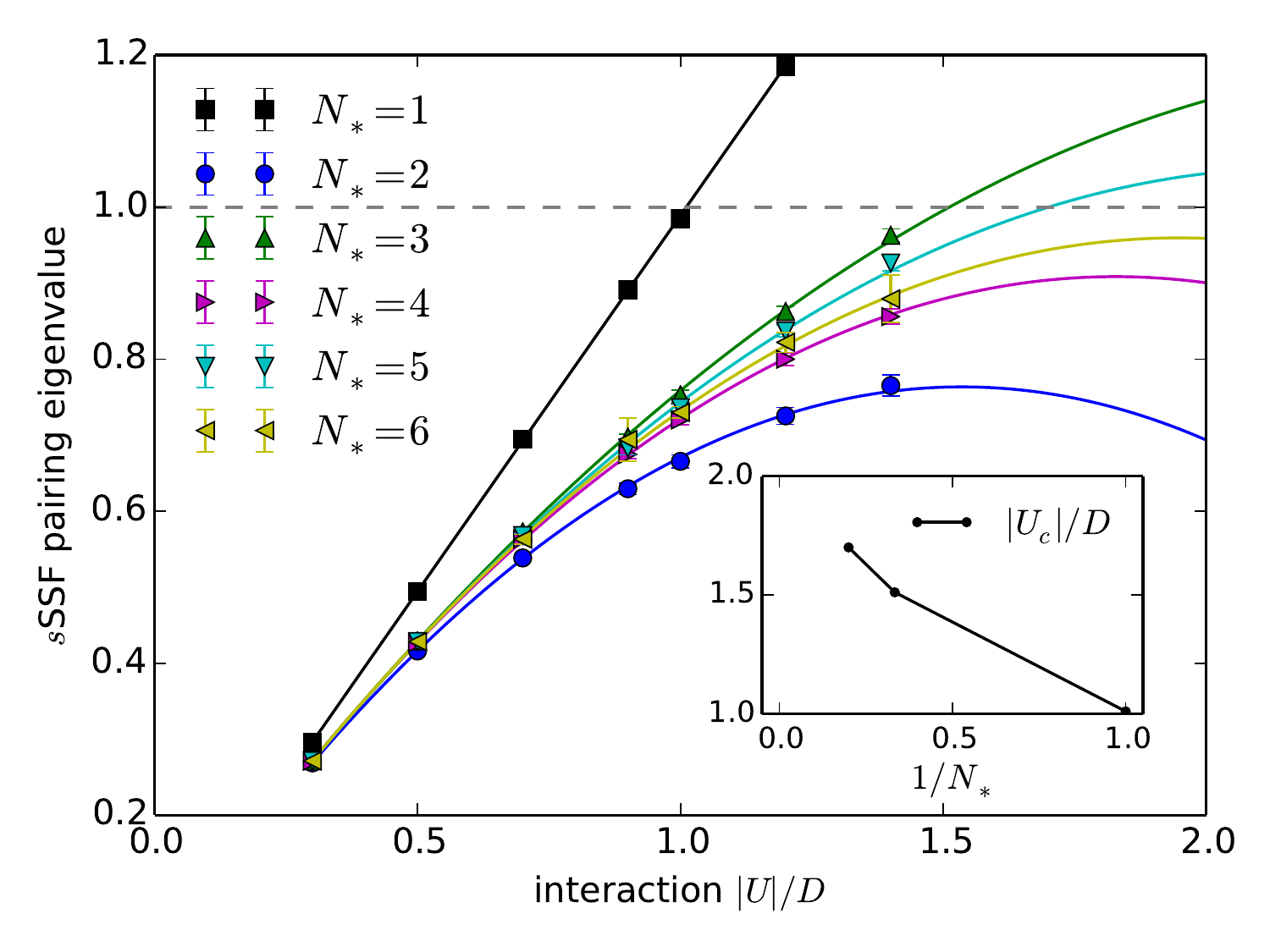}
	\includegraphics[width=.96 \columnwidth]{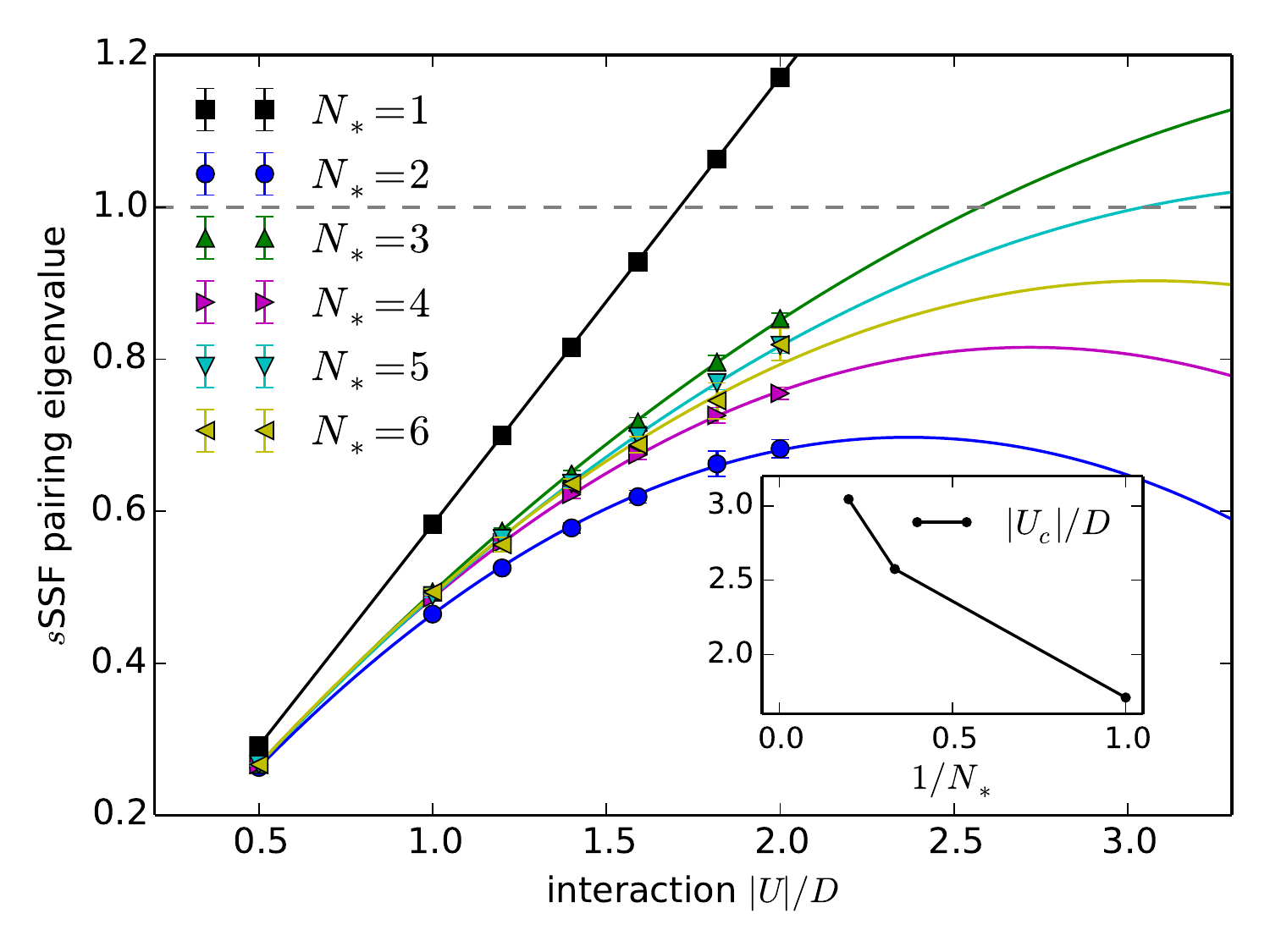}
	\caption{Saturated eigenvalues in the $s$SSF channel for constant anisotropy $\alpha=0.5$ \textit{(top)} and $\alpha=0.9$ \textit{(bottom)}. We find lower bounds $|U_c|(\alpha=0.5) \geq 1.5 D$ and $|U_c|(\alpha=0.9) \geq 2.5 D$ for the onset of a BCS instability.
	}
	\label{fig:swave-onset-U}
\end{figure}

Finally, we present results of additional DiagMC simulations used to characterize the phase diagram introduced in the main text.
For the isotropic limit (Fig.~\ref{fig:lambda_T_U1_alpha0}) we find a $s$SSF transition temperature which is in quantitative agreement with the transition temperatures found by Paiva \etal \cite{paiva2004} with the Determinant Monte Carlo algorithm, which does not suffer from a sign problem for attractive interactions in the presence of spin inversion symmetry.
\begin{figure}
	\includegraphics[width=.95 \columnwidth]{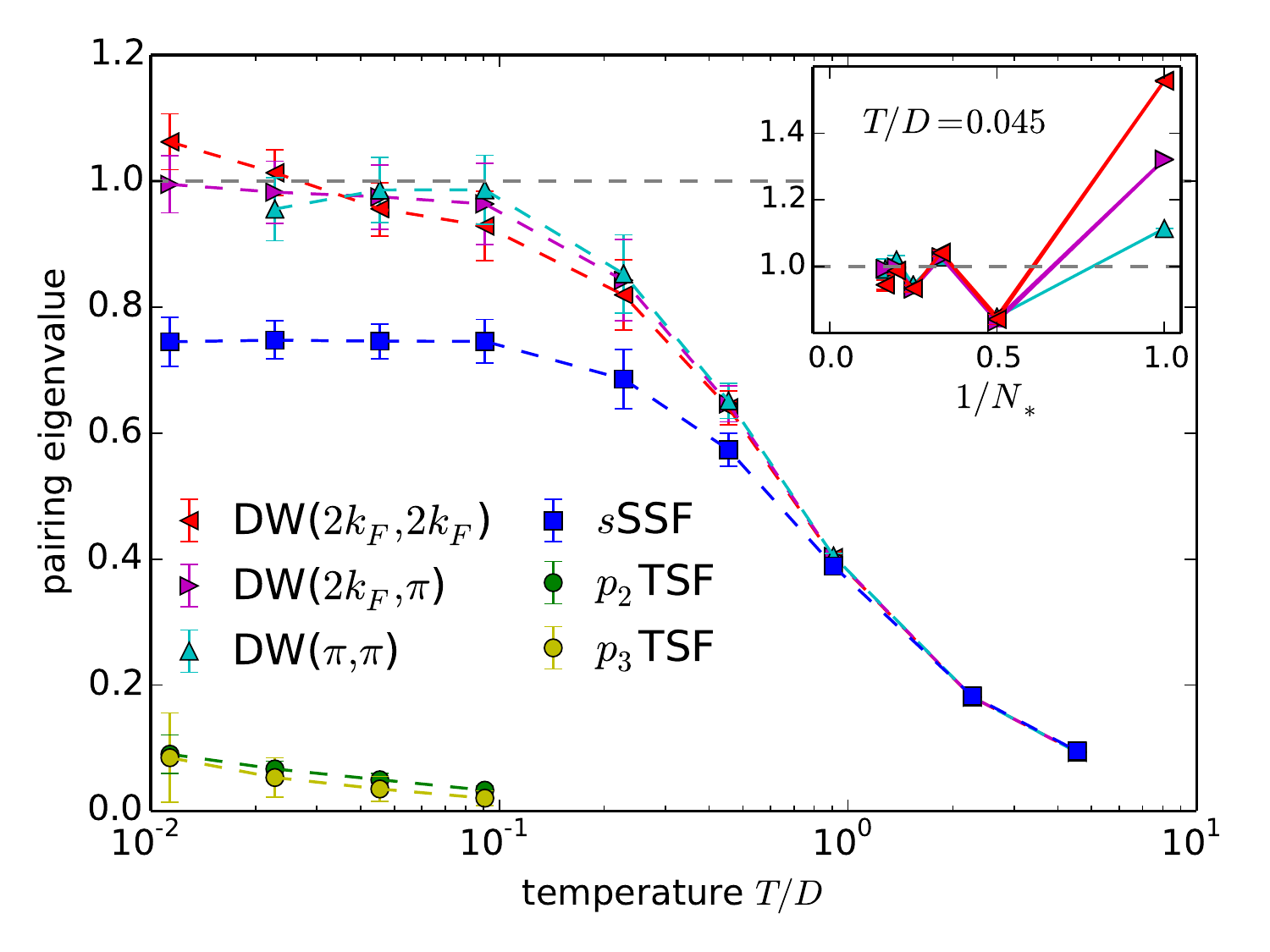}
	\caption{Leading eigenvalues for strong anisotropy $\alpha=0.9$ and interaction $|U|=1.8 D$. There is a close competition between density waves with different ordering momenta. Their error bars are dominated by uncertainties in the $N_* \to \infty$ extrapolation. The inset shows convergence of the different DW eigenvalues with diagram order $N_*$ for one temperature.}
	\label{fig:lambda_T_U1.8_alpha0.9}
\end{figure}
As soon as spin symmetry is broken by a finite anisotropy, the Cooper instability at weak coupling is removed but a sizeable interaction may still drive a transition at higher temperatures where the FS mismatch is less relevant. We observe the generic feature that the pairing eigenvalues rise quickly with decreasing temperature until a characteristic temperature $T^* \ll D$ where the mismatch is resolved and the eigenvalues saturate to a temperature-independent plateau.
Varying the interaction or anisotropy changes this characteristic temperature and, more importantly, the height of the low-temperature plateau, leading to a very sharp drop of the transition temperature to zero when the saturated eigenvalue drops below unity.
Instead of trying to resolve this extremely steep $T_c$ dependence we monitor the saturated eigenvalue well below $T^*$ while changing $\alpha$ or $U$ in order to determine the onset of the $s$SSF instability.
Fig.~\ref{fig:swave-onset-alpha} shows this procedure for determining the critical anisotropy $\alpha_c$ at weak and intermediate interaction $|U|=D/2, D$.
\begin{figure}
	\includegraphics[width=.91 \columnwidth]{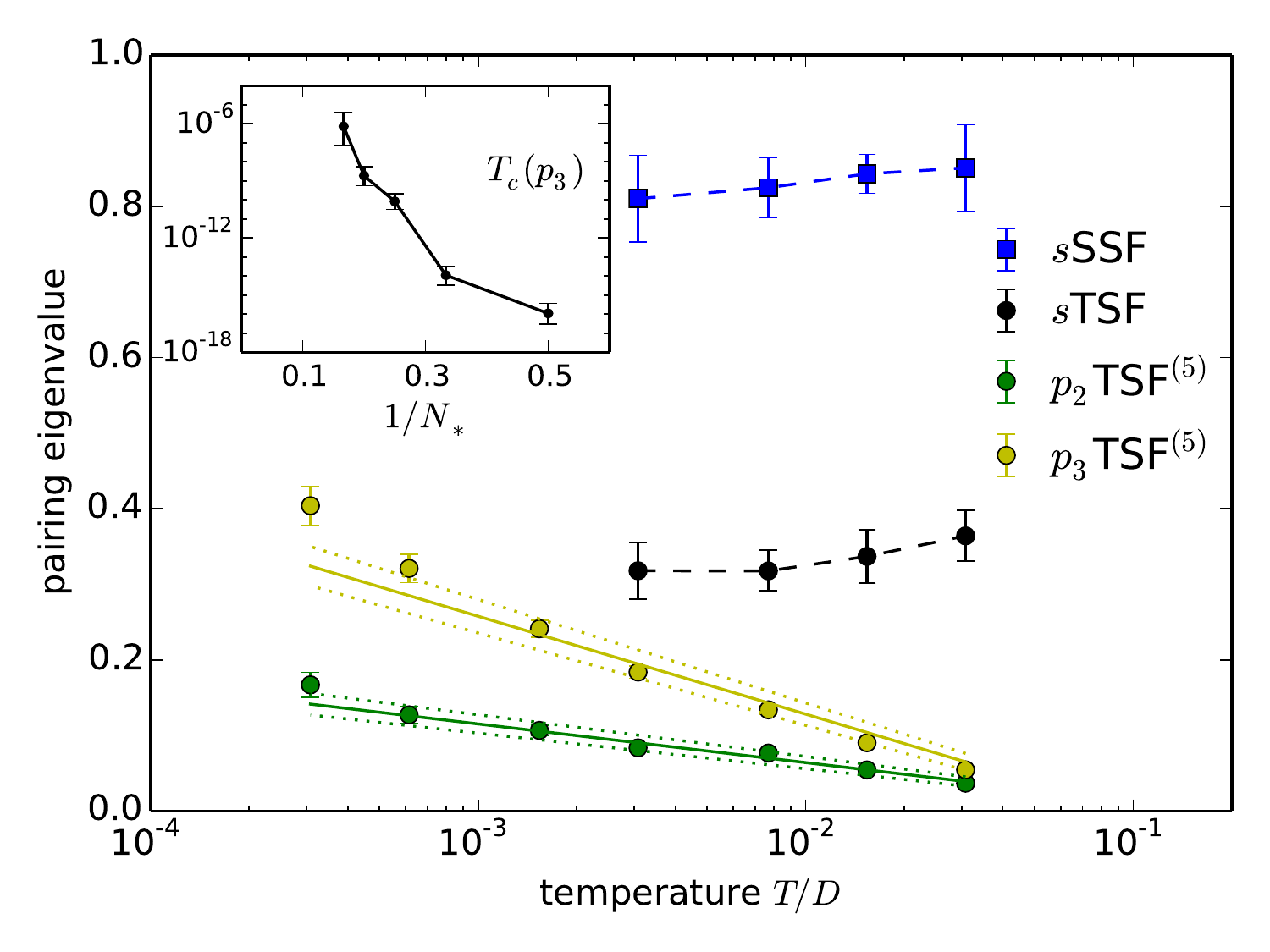}
	\includegraphics[width=.93 \columnwidth]{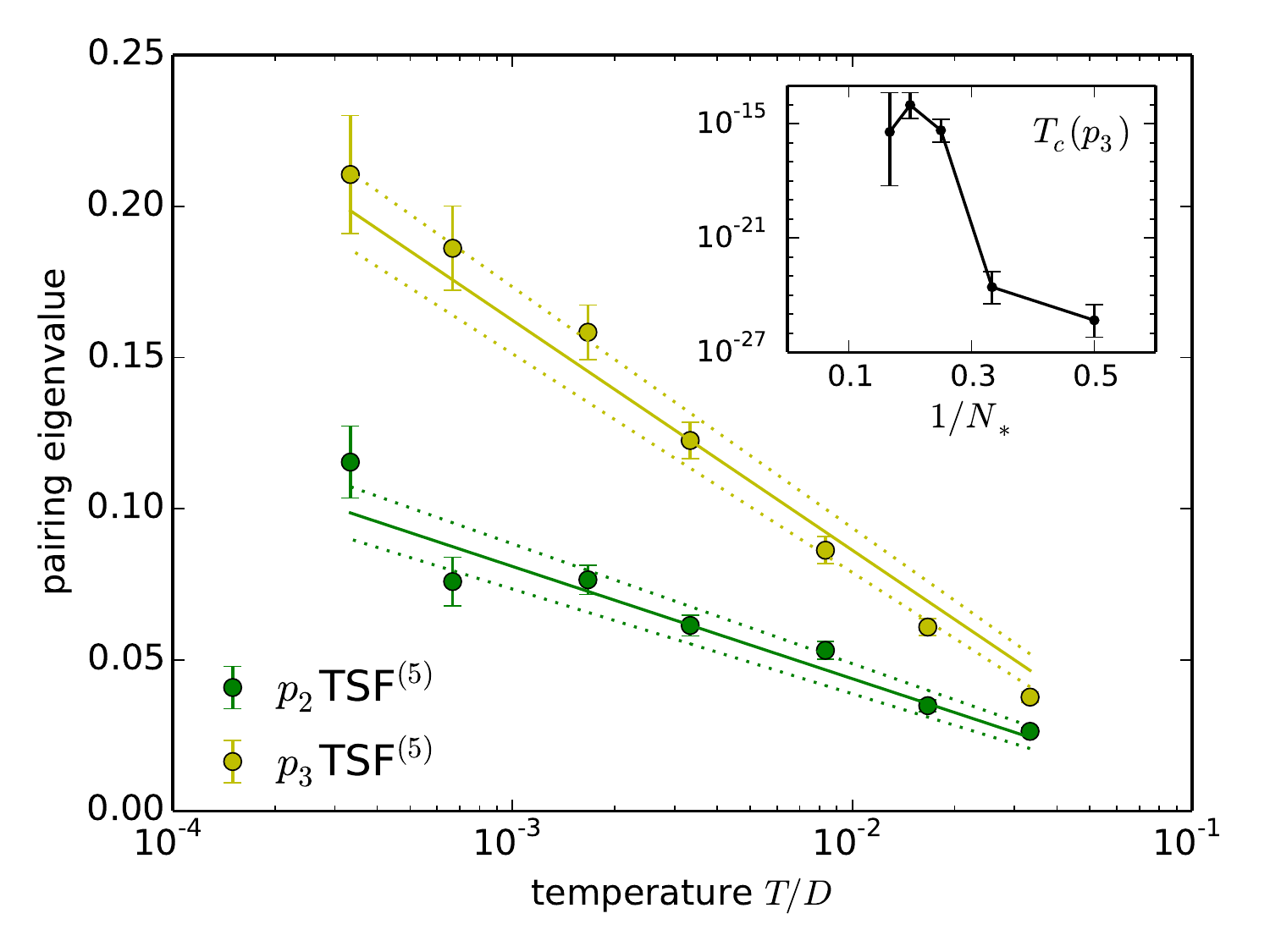}
	\caption{Superconducting eigenvalues for intermediate anisotropy $\alpha=0.375$ \textit{(top)} and $\alpha=0.5$ \textit{(bottom)}. Due to slow convergence $p$-wave eigenvalues in the main plots are for fixed cut-off $N_*=5$. Insets show how the transition temperature in the leading $p_3$ channel changes with order.
	Solid lines through $p$-wave data points are the low temperature asymptotics with only a constant offset fitted to the finite $T$ data points. Dotted lines mark the uncertainty in the slopes due to stochastic errors.}
	\label{fig:lambda_T_U1_alpha0.5}
\end{figure}
At larger interactions the diagrammatic series converge too slowly for a reliable localization of the onset. Varying the interaction for constant anisotropy, shown in Fig.~\ref{fig:swave-onset-U}, we can only give lower bounds on the critical interaction $|U_c|$ for strong anisotropy $\alpha \gtrsim 0.5$.
If there is any instability in the $s$SSF channel at these strong anisotropies it is most probably preempted by a transition in the particle-hole channel as we show explicitly for one point at strong interaction and anisotropy (Fig.~\ref{fig:lambda_T_U1.8_alpha0.9}).
While this case shows a close competition of density waves with different momenta, we expect that smaller anisotropies would disfavor momenta related to the $2k_F$ nesting around extreme anisotropy whereas $\vec{Q}=(\pi,\pi)$ order should be less sensitive to a change of anisotropy. Therefore we suspect that nearest-neighbor checkerboard order may dominate for a large range of anisotropies at strong interaction $|U| \gtrsim 1.8 D$. However, arguments based on FS matching become of course increasingly moot as the strong coupling limit is approached.
While the triplet superfluids dominating the phase diagram at intermediate anisotropy are arguably the most interesting phases they are also the most challenging to access numerically due to their exponentially low transition temperatures at moderate interaction strength and their first appearing at second order in the diagrammatic series.
The highest transition temperatures are to be expected at anisotropies close to the FS topology change $\alpha^*$ and at large interaction strength, \ie close the onset of $s$SSF or DW instabilities.
In these regimes the $p$-wave channels converge only slowly with diagram order, preventing reliable extrapolations.
The qualitative picture, however, is very consistent and independent of the chosen cut-off order $N_*$.
At the anisotropies studied in Fig.~\ref{fig:lambda_T_U1_alpha0.5} the $p_3$ symmetry clearly dominates over $p_2$ at all orders in accordance with the weak coupling data.
Slopes of the finite temperature eigenvalues in $\ln T$, which are calculated taking the full momentum and frequency structure of pair propagator and vertex into account, agree well with Fermi liquid extrapolations.
Higher order diagrams for the irreducible vertex consistently increase $\lambda_{FS}$, resulting in an exponential growth of $T_c$.
Given this strong trend towards larger $T_c$ visible in the insets of Fig.~\ref{fig:lambda_T_U1_alpha0.5} and the onset of the competing $s$SSF phase being only at significantly larger interaction strength than the $|U|=D$ considered here it is plausible that these exotic triplet superfluids could be observable at temperatures several orders of magnitude higher than the finite-order $T_c$ estimates shown here.


\bibliography{refs}{}

\end{document}